\documentclass[journal]{IEEEtran}
\usepackage[table,xcdraw]{xcolor}
\usepackage{amsmath,amssymb,amsfonts}
\usepackage{algorithmic}
\usepackage{graphicx}
\usepackage{textcomp}
\usepackage[caption=false,font=footnotesize]{subfig}
\usepackage{etoolbox}
\makeatletter
\usepackage{soul}
\usepackage{array}
\usepackage[ruled, linesnumbered, vlined]{algorithm2e}
\usepackage{textcomp}
\usepackage{multirow}
\usepackage{bm}
\usepackage{booktabs}
\usepackage{ntheorem}
\usepackage{orcidlink} 
\hypersetup{hidelinks}
\usepackage{mathrsfs}
\usepackage{mathtools}
\usepackage{epstopdf}
\usepackage{cases}
\def\BibTeX{{\rm B\kern-.05em{\sc i\kern-.025em b}\kern-.08em
		T\kern-.1667em\lower.7ex\hbox{E}\kern-.125emX}}
\usepackage{cases}
\theoremseparator{.}

\theorembodyfont{}

\definecolor{color}{rgb}{0, 0, 0}
\definecolor{color_blue}{rgb}{0, 0, 0}
\begin{document}

\title{Wireless Laser Power Transfer for Low-altitude Uncrewed Aerial Vehicle-assisted Internet of Things: Paradigms, Challenges, and Solutions}

\author{Chengzhen~Li,
        Likun~Zhang, 
        Chuang~Zhang*,
        Jiahui~Li,
        Changyuan~Zhao,\\
        Ruichen~Zhang,
        and Geng~Sun,~\IEEEmembership{Senior Member,~IEEE}

        \IEEEcompsocitemizethanks{
            \IEEEcompsocthanksitem Chengzhen Li, Likun Zhang and Jiahui Li are with the College of Computer Science and Technology, Jilin University, Changchun 130012, China (e-mails: zafro5791@gmail.com; zhanglk23@mails.jlu.edu.cn; lijiahui@jlu.edu.cn). 
            \IEEEcompsocthanksitem Chuang Zhang is with the College of Computer Science and Technology, Jilin University, Changchun 130012, China, and also with the Singapore University of Technology and Design, Singapore 487372 (e-mail: chuangzhang1999@gmail.com).
            \IEEEcompsocthanksitem Changyuan Zhao and Ruichen Zhang are with the College of Computing and Data Science, Nanyang Technological University, Singapore 639798 (e-mails: zhao0441@e.ntu.edu.sg; ruichen.zhang@ntu.edu.cn). 
            \IEEEcompsocthanksitem Geng Sun is with the College of Computer Science and Technology, Key Laboratory of Symbolic Computation and Knowledge Engineering of Ministry of Education, Jilin University, Changchun 130012, China, and also with the College of Computing and Data Science, Nanyang Technological University, Singapore 639798 (e-mail: sungeng@jlu.edu.cn).}
            \thanks{\textit{(Corresponding author: Chuang Zhang.)}}

}

\markboth{Journal of \LaTeX\ Class Files,~Vol.~X, No.~X, July~2025}%
{Shell \MakeLowercase{\textit{et al.}}: Bare Demo of IEEEtran.cls for Computer Society Journals}

\IEEEtitleabstractindextext{
	\begin{abstract}		
		Low-altitude uncrewed aerial vehicles (UAVs) have become integral enablers for the Internet of Things (IoT) by offering enhanced coverage, improved connectivity and access to remote areas. A critical challenge limiting their operational capacity lies in the energy constraints of both aerial platforms and ground-based sensors. This paper explores WLPT as a transformative solution for sustainable energy provisioning in UAV-assisted IoT networks. We first systematically investigate the fundamental principles of WLPT and analysis the comparative advantages. Then, we introduce three operational paradigms for system integration, identify key challenges, and discuss corresponding potential solutions. In  case study, we propose a multi-agent reinforcement learning framework to address the coordination and optimization challenges in WLPT-enabled UAV-assisted IoT data collection. {\color{color_blue}{Simulation results demonstrate that our framework achieves up to 15.1\% reduction in peak AoI compared to conventional multi-agent deep reinforcement learning (MADRL) methods.}} Finally, we discuss some future directions.
	\end{abstract}
	
	\begin{IEEEkeywords}
		Internet of things, wireless power transfer, uncrewed aerial vehicle, low-altitude wireless networks, multi-agent system
\end{IEEEkeywords}}

\maketitle
\IEEEdisplaynontitleabstractindextext
\IEEEpeerreviewmaketitle

\section{Introduction}
\label{Section: Introduction}

\par \IEEEPARstart{D}{ue} to flexible operational capabilities, the integration of low-altitude uncrewed aerial vehicles (UAVs) into the Internet of things (IoT) applications marks a notable progression in wireless sensor networks, which facilitates dynamic data acquisition and enhances mobile sensing functionalities \cite{Wang2025}. Specifically, compared with conventional ground-based IoT networks, low-altitude UAV-assisted IoT networks achieve extended coverage, superior line-of-sight connectivity, and access to remote or otherwise unreachable areas \cite{He2024}. These advantages make them highly suitable for applications such as precision agriculture, emergency response operations and smart city infrastructure management \cite{Yuan2025}. However, the effectiveness of low-altitude UAV-assisted IoT networks is fundamentally constrained by energy limitations across both low-altitude aerial platforms and terrestrial IoT sensors. {\color{color_blue}{For example, consumer drones produced by DJI Technology Co., Ltd. usually have a flight duration of 30 to 50 minutes\footnote{{\color{color_blue}{https://www.dji.com/sg/products/comparison-consumer-drones}}} and some multi-sensor LoRa nodes operating solely on battery power without any energy harvesting assistance can last for only about 46 days\footnote{{\color{color_blue}{https://arxiv.org/abs/2506.03203}}} in low-power mode.}} Conventional power supply methods, such as manual battery replacement or wired charging, are often logistically impractical for IoT devices distributed across large or inaccessible areas, and entirely infeasible for UAVs during flight operations \cite{Liu2025}. Therefore, the urgent demand arises for innovative energy provisioning technologies capable of overcoming these persistent constraints and enabling the sustainable and reliable operation of low-altitude UAV-assisted IoT systems.

\par Wireless power transfer (WPT) technologies have emerged as a promising solution to address these energy constraints \cite{Ren2025}. Unlike conventional wired charging, WPT provides contactless energy delivery based on different physical mechanisms. The advantages of WPT lie in continuous power supply to mobile devices, operation in inaccessible or hazardous environments, and reduced maintenance demands. Nonetheless, in low-altitude UAV-assisted IoT applications, conventional WPT approaches exhibit inherent limitations. In particular, inductive and capacitive power transfer methods are restricted to very short operational distances and require precise alignment between transmitter and receiver \cite{Wang2023}. Moreover, although microwave and radio frequency-based systems are capable of long-range energy delivery, their inherently omnidirectional radiation patterns result in considerable energy inefficiency, introduce potential electromagnetic interference with coexisting communication systems, and are further constrained by stringent regulatory restrictions on transmission power \cite{Xie2025}.

\begin{figure*}
    \centering
    \includegraphics[width=\linewidth]{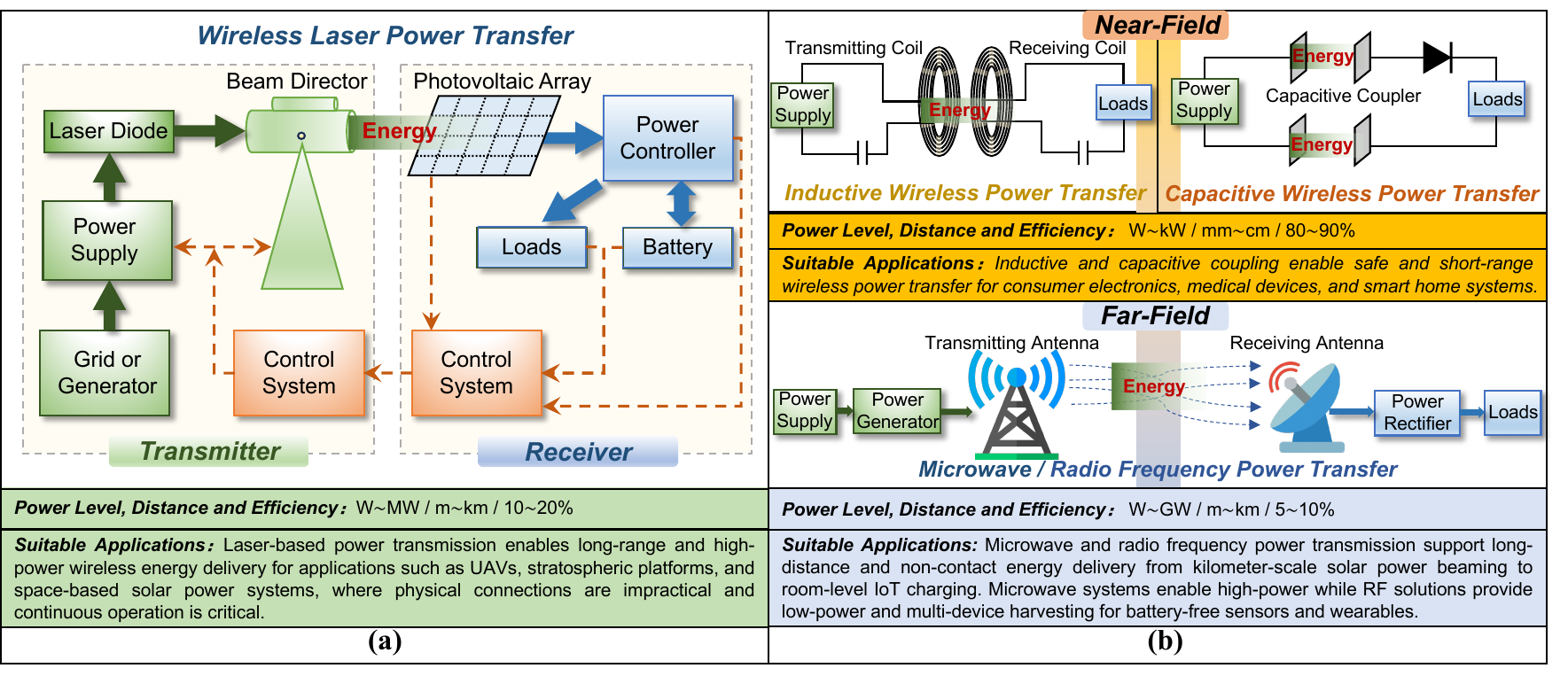}
    \caption{{\color{color_blue}{Principles, system architectures and application domains of major WPT technologies.}}}
    \label{Fig: Mag_Fig1}
\end{figure*}

\par In contrast, wireless laser power transfer (WLPT) represents a transformative approach that effectively overcomes these limitations while preserving the fundamental advantages of WPT technology \cite{Ahmadi2025}. Specifically, WLPT utilizes highly collimated laser beams to achieve efficient energy transmission over long distances, substantially exceeding the range limitations of near-field coupling methods. Moreover, the exceptional directional characteristics of laser beams typically maintain low divergence angles, thus permitting precise spatial control of energy delivery with minimal radiation loss to the surrounding environment. Furthermore, operation in the optical spectrum eliminates electromagnetic interference concerns with existing communication systems while offering enhanced security through the difficulty of intercepting narrowly focused energy beams. {\color{color_blue}{From the perspective of prototypes, the 2019 field demonstration conducted by the U.S. Naval Research Laboratory represented an important milestone, achieving 400 W laser power transmission across 325 m using a 2 kW laser transmitter coupled with a wavelength-optimized photovoltaic receiver\footnote{{\color{color_blue}{https://www.nrl.navy.mil/Media/News/Article/2504007/researchers-transmit-energy-with-laser-in-historic-power-beaming-demonstration}}}.}}

\par Motivated by this, we present a comprehensive investigation of WLPT for low-altitude UAV-assisted IoT systems by examining fundamental principles, operational paradigms and practical implementation challenges. The main contributions of this paper are summarized as follows.

\begin{itemize}
    \item We systematically analyze the fundamental principles of WLPT and its comparative advantages for aerial applications, thereby demonstrating its superior performance over conventional WPT methods in transmission range, directional precision and interference resistance.
    \item We propose three innovative operational paradigms in WLPT for UAV-assisted IoT, identify major implementation challenges in dynamic operational environments, and also discuss the corresponding potential solutions.
    \item We design a novel multi-agent proximal policy optimization with temporal memory and multi-agent coordination framework (MAPPO-TM) and introduce a case study on WLPT for UAV-assisted IoT data collection. Simulation results show the effectiveness of our framework in balancing energy sustainability and data freshness.
\end{itemize}
\section{Principles and Advantages of WLPT}
\label{Section: Principles and Advantages of WLPT}

\subsection{Principles of WLPT}
\label{Subsection: Principles of WLPT}
\par Based on the photoelectric effect principle, WLPT exploits laser beams as a carrier for energy transmission. As shown in Fig. \ref{Fig: Mag_Fig1}(a), a typical WLPT system consists of a laser transmitting device that converts electrical energy into an optical beam, and a photovoltaic-based receiving device that reconverts the incident laser power back into electricity \cite{Jin2019}. Specifically, the principles of both the power transmitter and the power receiver are as follows:
\begin{itemize}
    \item \textbf{\textit{Power Transmitter Side}:} 
    The initial electrical energy of the WLPT system is provided by a grid or generator and then regulated by a laser power supply before being fed into the laser diode. The laser diode subsequently converts the input electrical power into coherent laser radiation with a narrow divergence. Next, the generated laser beam is processed by the laser beam director (LBD), which performs collimation and steering to ensure accurate pointing toward the receiver. Moreover, a control system continuously monitors the output power and beam direction, and dynamically adjusts the laser operation based on feedback from the receiver.
    
    \item \textbf{\textit{Power Receiver Side}:} 
    In the context of the receiving device, the incident laser is first captured by a photovoltaic array, thereby converting optical energy back into direct current electricity. Then, the harvested electricity is regulated by a power controller, which distributes the energy either directly to the loads for real-time operation or to the battery for storage. In addition, a control system that matches the transmit side supervises the power regulation process and provides feedback to the transmitter.
\end{itemize}

\subsection{Advantages over Other WPT Methods}
\label{Subsection: Advantages over Other WPT Methods}

\par The other WPT methods can be categorized into four primary approaches according to distinct physical principles, which includes inductive wireless power transfer (IWPT), capacitive wireless power transfer (CWPT), microwave power transfer (MPT) and radio frequency power transfer (RFPT). The principles and comparisons among these WPT methods are illustrated in Fig. \ref{Fig: Mag_Fig1}(b). Building upon this comparison, the key advantages of WLPT can be summarized as follows.

\begin{itemize}
    \item \textbf{\textit{Extended Transmission Distance}:} WLPT fundamentally transcends the distance limitations inherent in conventional WPT methods through the low-divergence characteristics and minimal atmospheric propagation losses of laser. While near-field coupling methods such as IWPT and CWPT are constrained to operational ranges within centimeters due to the exponential decay of electromagnetic fields, WLPT enables power transmission across distances ranging from meters to potentially kilometers. This extended range capability is achieved without the inverse-square power density degradation that fundamentally limits far-field radiative approaches including MPT and RFPT.
    
    \item \textbf{\textit{Superior Spectral Characteristics}:} The utilization of optical frequency spectrum endows WLPT with distinctive advantages over conventional electromagnetic-based WPT systems. Operating in the optical domain eliminates the electromagnetic interference concerns associated with radio frequency-based methods such as IWPT, CWPT and RFPT, which function within congested spectral bands and may disrupt sensitive electronic equipment. Furthermore, WLPT circumvents the regulatory constraints and biological safety concerns related to microwave radiation exposure present in MPT systems, while providing access to a broad optical spectrum.
    
    \item \textbf{\textit{Enhanced Beam Directionality}:} WLPT demonstrates superior spatial power control through the inherent collimation properties of laser radiation and advanced optical beam management techniques. The highly directional nature of laser beams enables precise spatial power delivery with low beam divergence angles, fundamentally eliminating the omnidirectional power loss characteristic of far-field methods such as MPT and RFPT. Additionally, WLPT incorporates sophisticated beam steering and adaptive focusing capabilities that surpass the stringent physical alignment requirements of near-field coupling systems including IWPT and CWPT, thus enabling dynamic power delivery to mobile or misaligned receivers.
\end{itemize}



\section{WLPT for Low-altitude UAV-assisted IoT}
\label{Section: WLPT for Low-altitude UAV-assisted IoT}
\begin{figure*}
    \centering
    \includegraphics[width=\linewidth]{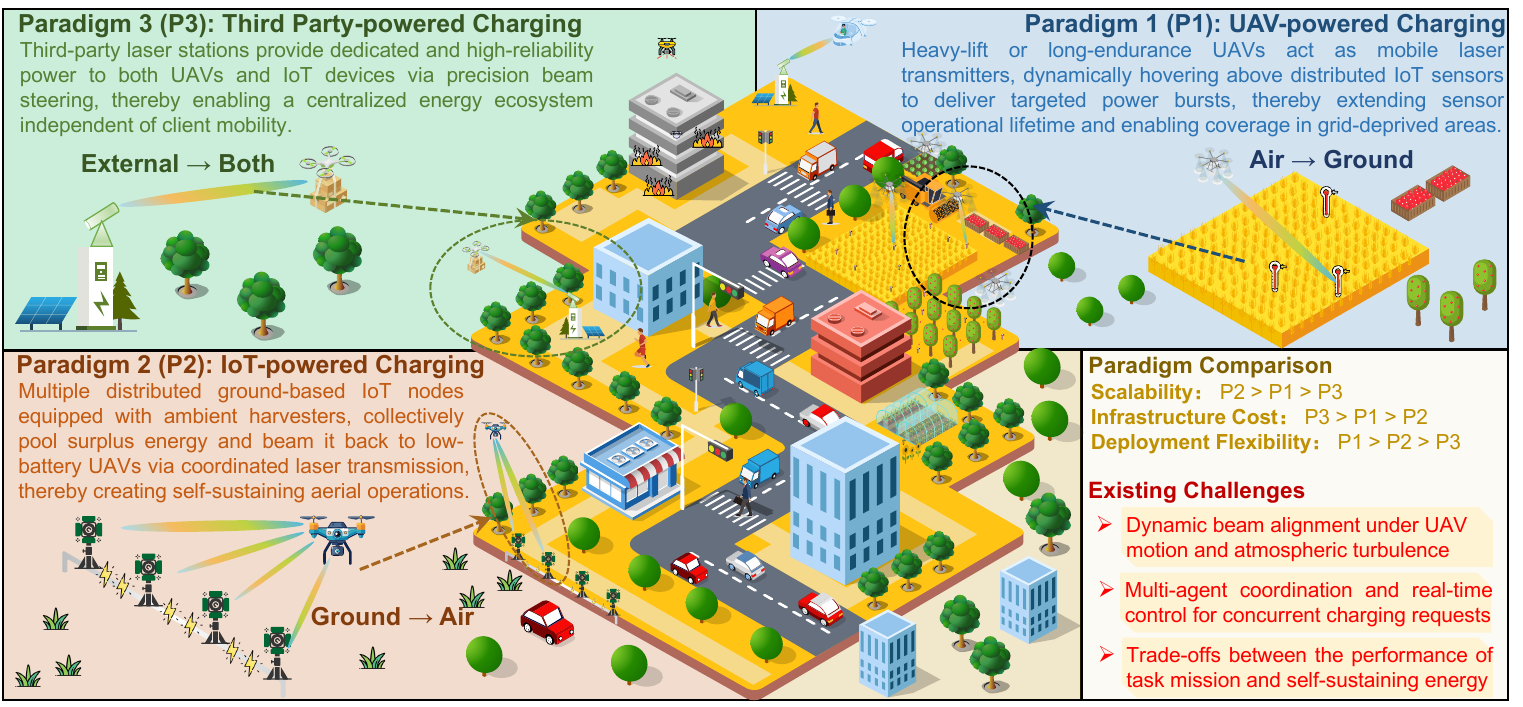}
    \caption{{\color{color_blue}{Paradigms, comparison and challenges of WLPT for low-altitude UAV-assisted IoT.}}}
    \label{Fig: Mag_Fig2}
\end{figure*}

\subsection{Paradigms of WLPT in Low-altitude UAV-assisted IoT}
\label{Subsection: Paradigms of WLPT in Low-altitude UAV-assisted IoT}

\par WLPT creates unprecedented opportunities for revolutionizing power delivery in low-altitude UAV-assisted IoT systems. Depending on different sources of the transferred energy, we consider the following important paradigms to exploit WLPT for low-altitude UAV-assisted IoT, as illustrated in Fig. \ref{Fig: Mag_Fig2}.

\begin{itemize}
    \item \textbf{\textit{UAV-powered IoT Charging}:} In this paradigm, UAVs with large battery capacities, such as heavy-lift rotorcraft, long-endurance fixed-wing platforms and hybrid VTOL aircraft, serve as aerial charging stations that supply wireless power to energy-constrained IoT devices through laser beams. Unlike static charging stations, these UAVs can flexibly adjust their positions to match the spatial distribution of IoT devices, thereby achieving timely and efficient energy replenishment \cite{Gupta2025}. By flying to device deployment areas and leveraging their altitude advantage, UAVs can establish efficient line-of-sight laser links that minimize energy leakage while their mobility enables them to sequentially serve multiple device clusters, thus enhancing the overall energy coverage of large-scale IoT systems. Therefore, UAV-powered IoT charging provides a promising solution for extending the lifespan and reliability of energy-constrained IoT networks.
    
    \item \textbf{\textit{IoT-powered UAV Charging}:} Conversely, the IoT-powered UAV charging paradigm harnesses the distributed energy harvesting capabilities of ground-based IoT networks to sustain aerial platforms with limited onboard power \cite{Yan2020}. This approach supports lightweight UAVs designed for surveillance, environmental sensing, or infrastructure inspection by enabling them to receive laser-based power replenishment from nearby IoT nodes. {\color{color_blue}{These nodes include infrastructure such as solar-powered environmental monitoring stations, fixed surveillance units, and agricultural gateways, which typically generate energy surplus beyond their own operational requirements. Moreover, their physical size can reasonably accommodate the additional hardware for laser transmission, unlike miniaturized sensors.}} Through coordinated laser beaming protocols, they dynamically pool harvested energy and transmit it to visiting UAVs. The dense spatial distribution of such IoT infrastructure ensures broad energy coverage across operational zones, while collaborative power sharing significantly extends UAV endurance and mission range. This symbiotic model thus offers a promising pathway toward self-sustaining aerial operations in remote or resource-limited environments.

    \item \textbf{\textit{Third Party-powered Charging}:} Different from the previous paradigms, third-party-powered charging introduces an extended energy service model in which neither UAVs nor IoT devices are expected to supply power to one another. Instead, both rely on independent external providers that may encompass ground-based laser stations, orbital satellites or stratospheric platforms \cite{Hassija2020}. These infrastructures are equipped with high-power laser transmitters, precision beam steering mechanisms and intelligent energy management protocols, thereby enabling them to serve heterogeneous clients simultaneously from compact sensor nodes to heavy-lift UAVs. Through coordinated network deployment and the integration of advanced beam tracking with adaptive power allocation algorithms, these providers can deliver targeted energy to UAVs in flight while ensuring uninterrupted power delivery to geographically dispersed IoT devices. Their specialized design guarantees high service reliability and energy transfer efficiency, while operational independence from client systems facilitates centralized optimization of power resources. As a result, third-party-powered charging establishes a scalable and dependable energy ecosystem, allowing aerial and terrestrial platforms to concentrate on mission-critical tasks while receiving assured power support from dedicated infrastructure. 
\end{itemize}

\par {\color{color_blue}{These paradigms differ in their primary energy source and power flow. While UAV-powered and IoT-powered paradigms enable bidirectional energy sharing between aerial and terrestrial nodes, the third-party paradigm provides a centralized power service, enhancing system flexibility and robustness. In dense deployments with multiple UAVs and laser-based transmitters, crosstalk may arise if beams overlap or wavelengths mismatch, potentially reducing photovoltaic conversion efficiency. Wavelength-division or time-division scheduling can mitigate these effects, ensuring each receiver operates within its optimal optical and spectral range and maintaining stable, efficient energy transfer.}}

\subsection{Applications of WLPT in Low-altitude UAV-assisted IoT}
\label{Subsection: Application of WLPT in Low-altitude UAV-assisted IoT}

\par To demonstrate the practical applicability of these WLPT paradigms, we examine three representative application scenarios across different operational environments.

\begin{itemize}
    \item \textbf{\textit{Precision Agriculture}:} In precision agriculture applications, UAVs equipped with high-capacity batteries serve as mobile charging platforms for distributed soil sensors, weather stations, and crop monitoring devices across vast farmlands. By dynamically navigating between different field sections, these aerial platforms deliver laser-based energy replenishment to IoT sensors that monitor soil moisture, nutrient levels and pest activity, thereby ensuring uninterrupted data collection even in remote zones. Conversely, during daylight hours, solar-powered agricultural IoT networks can collectively charge lightweight surveillance drones, which in turn enables extended missions for crop health assessment and livestock tracking. Moreover, third-party charging infrastructure, such as solar-powered laser stations positioned at field boundaries, can provide continuous energy support not only to UAVs conducting aerial surveys, but also to ground-based sensors performing long-term environmental monitoring, thus creating a symbiotic energy ecosystem across the agricultural landscape.

    \item \textbf{\textit{Emergency Response}:} During natural disasters or emergency situations, UAV-powered charging becomes critical for maintaining IoT sensor networks that monitor structural integrity, air quality and survivor detection systems in affected areas. Specifically, heavy-lift UAVs can rapidly deploy to disaster zones and provide emergency power to communication nodes, environmental sensors and rescue coordination devices, especially when traditional power infrastructure is compromised. Under such conditions, distributed IoT networks can support search-and-rescue drones through coordinated laser power transmission if ground-based energy harvesting systems remain partially operational, thereby extending their flight endurance for victim location and damage assessment. To further enhance resilience, emergency response centers can deploy portable third-party charging stations, establishing a decentralized yet reliable energy grid that simultaneously powers aerial reconnaissance platforms and ground-based emergency communication systems, thereby ensuring mission continuity amid chaos.

    \item \textbf{\textit{Smart City}:} Urban IoT deployments benefit significantly from UAV-assisted charging, particularly for traffic monitoring sensors, air quality detectors and smart lighting systems distributed throughout metropolitan areas. During peak monitoring periods, municipal UAVs can deliver targeted energy bursts to traffic management devices, while concurrently recharging parking sensors and environmental monitors in grid-constrained districts. In parallel, energy-efficient smart city networks equipped with building-integrated photovoltaics and kinetic energy harvesters can collectively power traffic surveillance drones and infrastructure inspection UAVs during high-demand intervals. Critically, city-wide third-party charging networks, including rooftop laser stations and integrated building energy systems, provide comprehensive power support not just to municipal UAV fleets, but also to the broader sensor fabric of the city ultimately guaranteeing the uninterrupted operation of essential urban services, from traffic flow optimization to pollution control.
\end{itemize}

\subsection{Challenges in WLPT for Low-altitude UAV-assisted IoT}
\label{Subsection: Challenges in WLPT for Low-altitude UAV-assisted IoT}

\par While the application scenarios demonstrate the promising potential of WLPT in UAV-assisted IoT systems, several critical challenges must be addressed to achieve practical implementation. 

\begin{itemize}
    \item \textbf{\textit{Laser Beam Alignment and Motion Tracking}:} The dynamic operations of low-altitude UAVs present significant challenges for maintaining precise laser beam alignment during wireless laser power transmission. These UAVs experience continuous positional variations due to wind disturbances, flight path adjustments and payload-induced oscillations, thus making it difficult to maintain the narrow beam focus required for efficient power transfer. Traditional stationary WLPT systems rely on fixed geometric relationships between transmitter and receiver, but UAV mobility introduces complex three-dimensional tracking requirements that must account for real-time position estimation errors, communication latencies and mechanical steering limitations of laser transmitters. Moreover, atmospheric effects such as turbulence, thermal gradients and particulate scattering can cause beam deviation and power density fluctuations, particularly in long-range power transmission scenarios. {\color{color_blue}{Furthermore, some obstacles may temporarily block the direct laser path, thus creating non-LoS conditions that interrupt power delivery.}}
    \item \textbf{\textit{Multi-agent Cooperation and Real-time Control}:} Effective coordination of multiple UAVs and IoT devices in a wireless laser power transmission network demands sophisticated multi-agent control algorithms. These algorithms must dynamically allocate resources, prevent aerial collisions and optimize energy distribution across the network. When multiple UAVs simultaneously request charging, real-time decision-making becomes especially challenging. In this case, intelligent scheduling mechanisms are required to prioritize access based on battery state, mission urgency and spatial constraints. Moreover, the inherent heterogeneity of the system with devices exhibiting diverse power demands, mobility behaviors and operational limits calls for adaptive coordination protocols. Such protocols must continuously reconfigure themselves in response to evolving network topology and fluctuating energy availability to maintain system-wide efficiency and stability.
    \item \textbf{\textit{Trade-offs between Energy and Task Performance}:} Implementing WLPT in UAV-assisted IoT systems introduces complex optimization challenges that involve balancing energy efficiency with operational performance. UAVs must distribute their limited onboard energy among propulsion, payload operation, communication and potential energy harvesting functions. This constraint gives rise to multi-objective optimization problems with inherently conflicting objectives. For example, in a scenario where a UAV periodically visits distributed IoT nodes to collect time-sensitive data while depending on laser-based charging stations for energy replenishment. In this setting, achieving low age of information (AoI) requires frequent visits and aggressive flight trajectories. Such behavior accelerates battery depletion and may necessitate premature or high-risk recharging. In contrast, energy-conserving strategies, such as reducing visit frequency or postponing charging, extend mission duration but result in increased data staleness and compromised timeliness.

\end{itemize}

\subsection{Solutions in WLPT for Low-altitude UAV-assisted IoT}
\label{Subsection: Solutions in WLPT for Low-altitude UAV-assisted IoT}

\par To overcome the critical challenges hindering the practical deployment of WLPT in low-altitude UAV-assisted IoT systems, the following possible solutions are envisioned.

\begin{itemize}

\item \textbf{\textit{Advanced Beam Steering and Tracking}:} {\color{color_blue}{Traditional control approaches, such as proportional-integral-derivative control or physical model-based methods, typically require an accurate system model and are often designed for static or mildly dynamic environments.}} To maintain efficient laser-enabled power transfer under UAV mobility, emerging WLPT systems in low-altitude UAV-assisted IoT can adopt vision-based deep learning models that fuse camera and inertial data to predict receiver motion and steer laser beams in real time. Unlike traditional geometry-based trackers, these learning-augmented approaches generalize across lighting, vibration and partial occlusion, thereby enabling plug-and-play deployment without labor-intensive calibration. This shift toward perception-driven control is redefining how aerial platforms interact with wireless energy infrastructure. For example, Xue \textit{et al.} \cite{Xue2025} proposed a similarity-guided layer-adaptive framework that dynamically prunes redundant vision transformer layers based on representation similarity, retaining only the most discriminative ones to achieve an optimal accuracy-speed trade-off for real-time UAV tracking. 

\begin{figure*}
    \centering
    \includegraphics[width=0.9\linewidth]{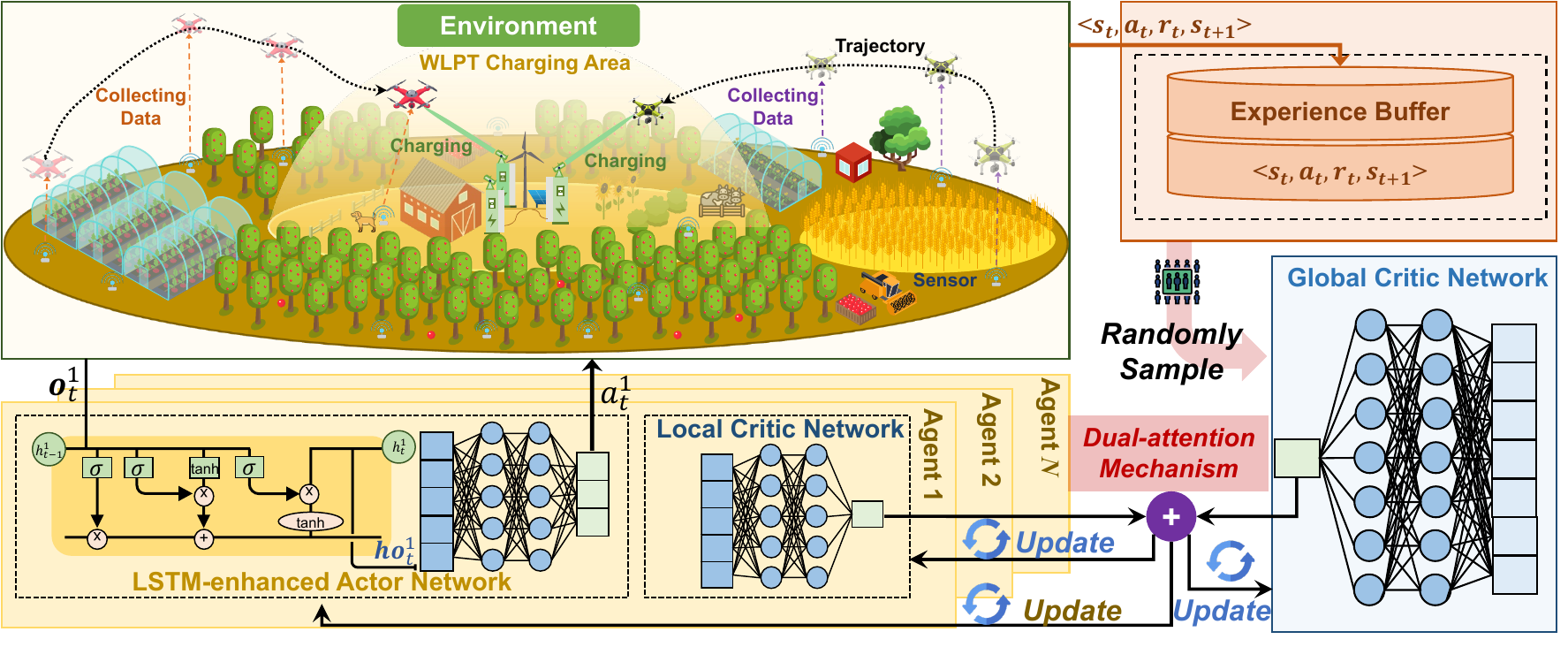}
    \caption{The framework of the proposed MAPPO-TM.}
    \label{Fig: Mag_Fig3}
\end{figure*}

{\color{color_blue}{\item \textbf{\textit{Fallback Strategies for Non-Line-of-Sight Scenarios}:} To mitigate the limitations caused by temporary obstructions of the direct laser path, two complementary strategies can be considered. Specifically, the first involves cooperative relay-assisted WLPT, in which additional low-altitude UAVs or ground nodes act as relay stations to forward laser energy from the source to the target. When the direct LoS path is blocked, these relay nodes provide alternative transmission paths, effectively extending coverage and reducing the likelihood of temporary power interruptions. Another strategy integrates WLPT with RFPT. When the direct laser path is unavailable, the system can temporarily switch to RFPT mode. Although radio frequency signals also experience attenuation due to obstacles, their longer wavelengths allow them to diffract around obstacles, reflect off surfaces, and partially penetrate materials, thereby maintaining a baseline level of power delivery until the direct laser link is restored.}}

\item \textbf{\textit{Lightweight Distributed Multi-agent Coordination}:} Efficient coordination in WLPT-enabled UAV-IoT networks can rely on multi-agent reinforcement learning (MARL), where UAVs learn to schedule charging requests and flight paths while minimizing interference and energy waste. Following the paradigm of centralized training and decentralized deployment, agents coordinate offline but execute autonomously online, ensuring seamless WLPT access without overloading communication channels. This intelligence layer is becoming essential for scaling laser-powered drone fleets in complex and dynamic environments. For instance, Dou \textit{et al.} \cite{Dou2025} presented a novel hybrid-action deep reinforcement learning framework, where the action decoder that decouples discrete and continuous actions was utilized in model training to address the challenge in learning the interdependency between UAV and mobile charger.

\item \textbf{\textit{Energy-Aware Multi-Dimension Joint Optimization}:} Sustaining long-term operations in WLPT-assisted systems requires joint optimization of flight, sensing and charging, which can be effectively addressed by model predictive control (MPC) and dynamic multi-objective optimization frameworks. These methods enable continuous trade-offs between mission responsiveness and energy consumption, ensuring UAVs receive laser power at the right time and place without compromising operational goals. By embedding energy awareness into the core of decision-making, WLPT systems gain the ability to operate resiliently even under resource fluctuations or task overloads. For example, in \cite{Zhang2025}, the authors studied a UAV-enabled WPT and data collection system for batteryless sensor networks, where a joint optimization of power allocation and trajectory was formulated to balance fair data collection and UAV energy efficiency.

\end{itemize}


\begin{figure*}[!t]
  \centering
  \subfloat[]{\includegraphics[width=0.325\linewidth]{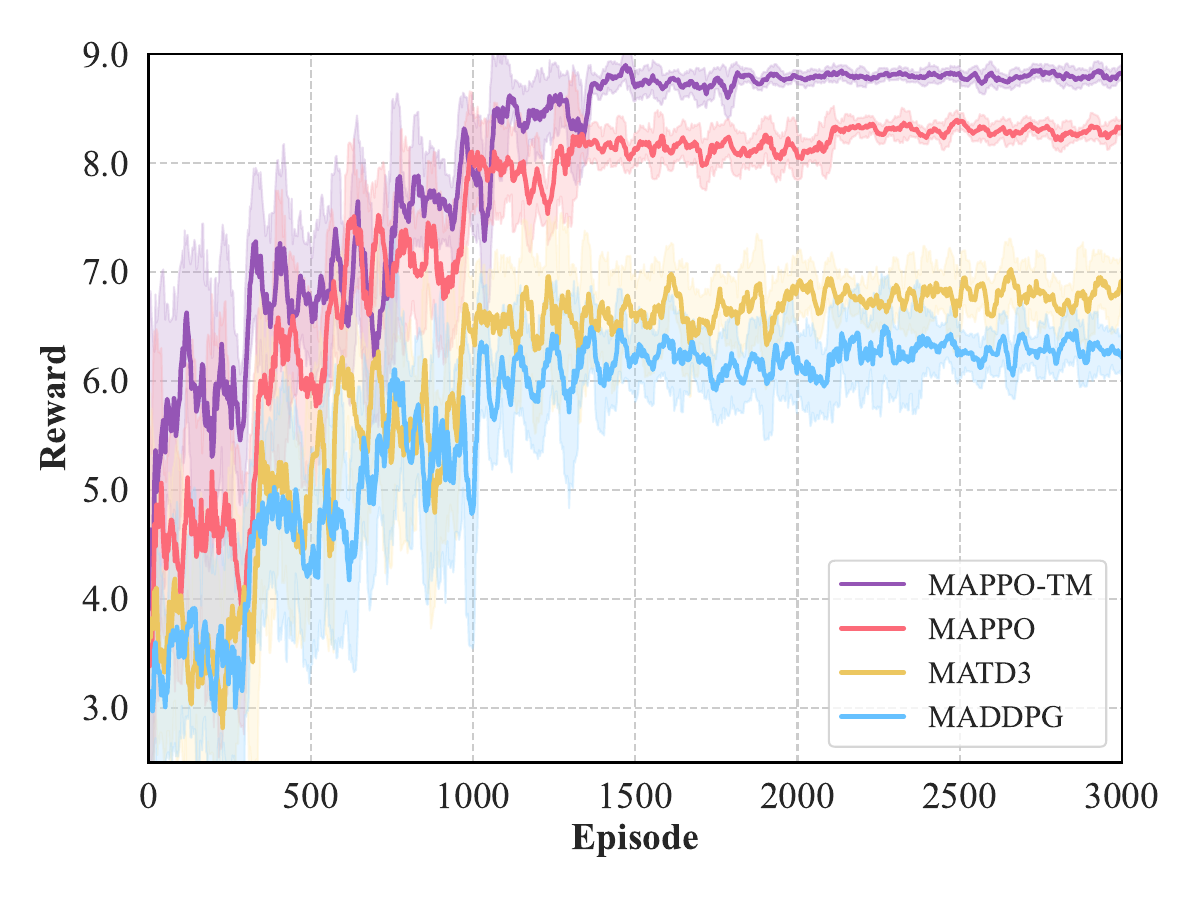}} \hfill
  \subfloat[]{\includegraphics[width=0.325\linewidth]{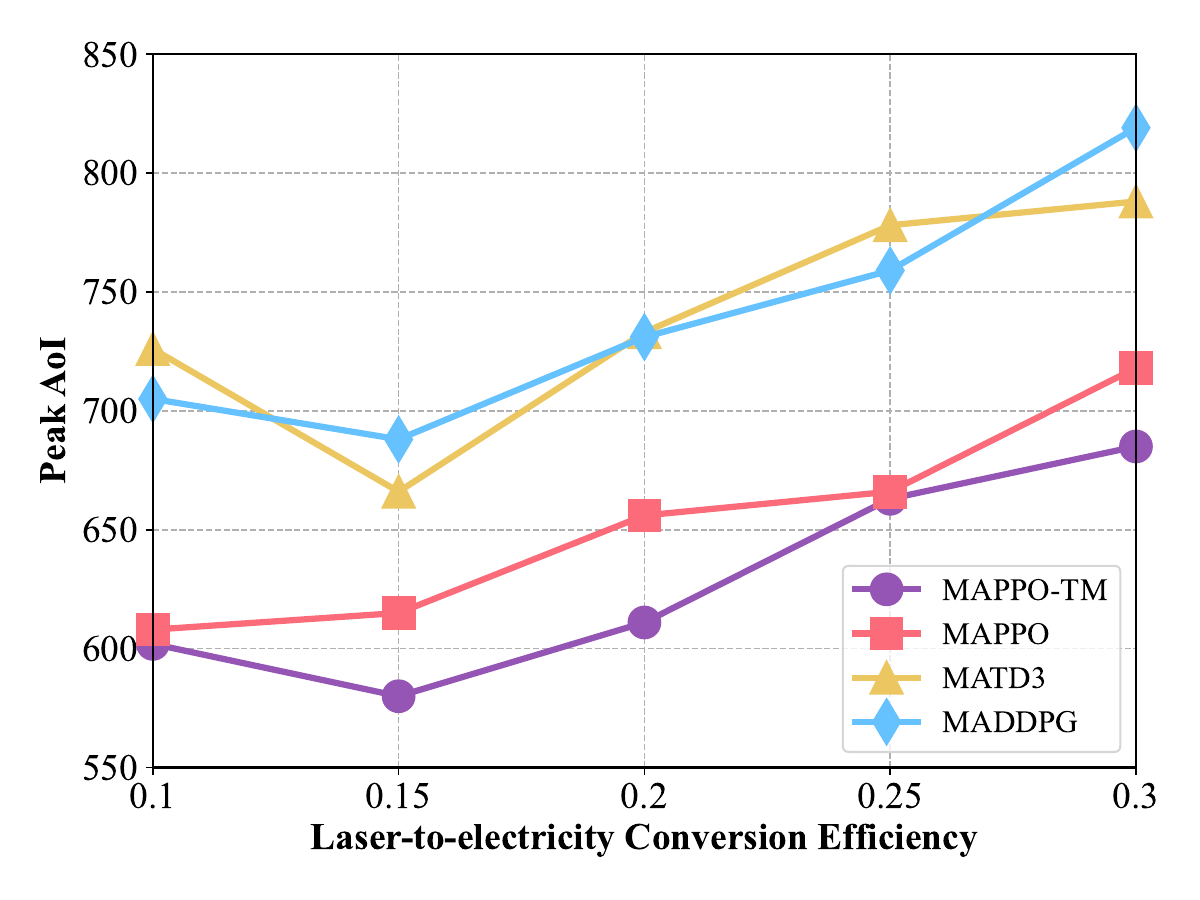}} \hfill
  \subfloat[]{\includegraphics[width=0.32\linewidth]{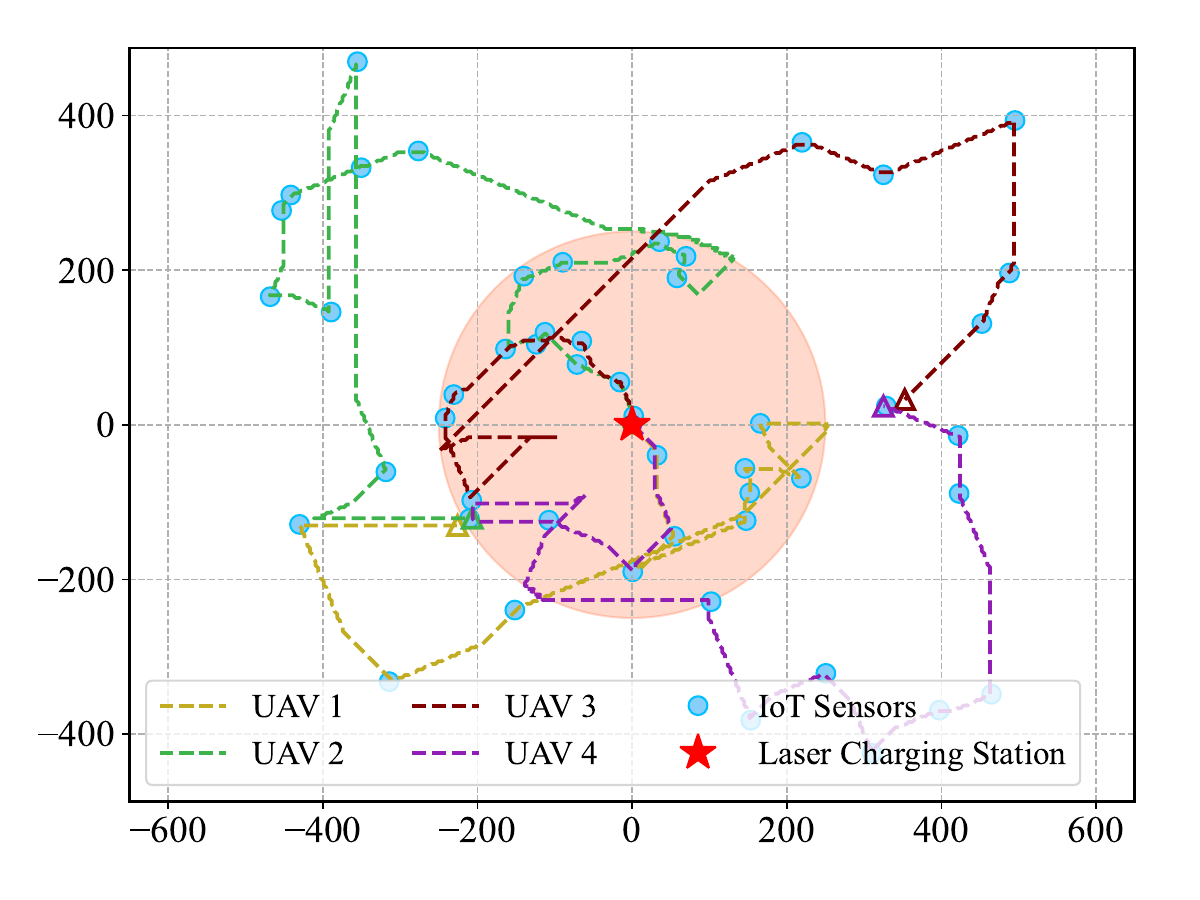}}
  \caption{Performance Comparison. (a) Reward training curves in terms of different algorithms. (b) Peak AoI comparison of different algorithms under various laser-to-electricity conversion efficiencies. (c) Trajectory of UAVs controlled by MAPPO-TM.}
  \label{Fig: Mag_Fig4}
\end{figure*}


\section{Case Study}
\subsection{Scenario Description}
\label{Subsection: Scenario Description}

\par We consider a $1000\times 1000$ m agriculture monitor area with $50$ static ground IoT sensors, deployed to continuously collect environmental data such as soil moisture, temperature, and crop health metrics. To ensure timely and reliable data acquisition across this expansive region, $4$ rotary-wing UAVs are deployed to periodically fly over the sensor network and collect sensed information. However, due to the limited onboard energy capacity of UAVs, we introduce a centralized WLPT infrastructure, in which a central charge station equipped with $10$ LBDs is arranged to form overlapping circular charging zones with a radius of $250$ m. Moreover, each UAV is equipped with a photovoltaic array capable of converting incident laser energy into electrical power with a laser-to-electricity efficiency $0.15$, operates at a constant altitude of $80$ m, and cruises at a speed of $5$ m/s\footnote{{\color{color_blue}{Altitude and cruise speed parameters of UAVs refer to Voyager: https://www.unmannedsystemstechnology.com/company/yellowscan}}}, thereby maintaining higher line-of-sight connectivity with both the IoT sensors and the central charge station. In this case, the UAVs face a fundamental trade-off, i.e., frequent data collection minimizes AoI but accelerates energy depletion, while conserving energy to extend mission duration risks excessive data staleness. 

\subsection{Proposed Framework}
\label{Subsection: Scenario Description}

\par To address the dynamic multi-agent coordination decision problem arising from the fundamental trade-off between minimizing AoI and preserving UAV operational endurance, we propose a novel MAPPO-TM framework. As shown in Fig. \ref{Fig: Mag_Fig3}, the core component of MAPPO-TM is listed as follows.

\par \textbf{\textit{Component 1:  Partially Observable Markov Decision Process}}. We reformulate the joint trajectory planning and charging strategy optimization problem as a partially observable Markov decision process. The state space captures the horizontal positions of all UAVs, their residual energy levels, and the AoI and status of all IoT sensor nodes. Each UAV observes only its own position, energy level, and the AoI of nearby IoT sensors within its communication range. The action space is discretized into eight cardinal and diagonal movement directions to ensure practical implementation.

\par \textbf{\textit{Component 2: Temporal Dependency Learning}}. Basic MAPPO processes only the current observation, which is insufficient for capturing historical patterns. To overcome this limitation, we embed a long short-term memory (LSTM) network into the actor network of each UAV. The LSTM takes the current local observation and the previous hidden state as inputs, thereby producing a new hidden state that encapsulates a compressed representation of the history.

\par \textbf{\textit{Component 3: Dual-Attention Coordination Mechanism}}. Basic MAPPO can lead to selfish behavior, where UAVs hoard energy or cluster around easily accessible nodes, thus neglecting distant ones and causing high AoI elsewhere. To foster cooperative behavior, we introduce a dual-attention mechanism within the centralized critic network. The critic computes two distinct value estimates: one representing the individual local performance and another capturing the global network-wide objective. These values are dynamically weighted by learnable parameters, allowing the system to adaptively balance individual survival instincts with cooperative mission goals.

\subsection{Simulation Results}
\label{Subsection: Simulation Results}

\par As shown in Fig. \ref{Fig: Mag_Fig4}(a), MAPPO-TM achieves the highest cumulative reward and demonstrates superior training stability compared to MAPPO, MATD3, and MADDPG, converging rapidly to a steady performance level. This is attributed to its LSTM-based temporal memory, which enables UAVs to leverage historical trajectory and energy patterns for more informed decision-making, while the dual-attention mechanism ensures balanced optimization of local energy management and global AoI minimization. The reduced variance in the reward curve of MAPPO-TM further confirms its robustness against environmental fluctuations and multi-agent interference. Moreover, Fig. \ref{Fig: Mag_Fig4}(b) illustrates the proposed MAPPO-TM algorithm achieves the lowest peak AoI across all benchmark algorithms under varying laser-to-electricity conversion efficiencies, demonstrating superior coordination and energy-aware decision-making. Furthermore, Fig. \ref{Fig: Mag_Fig4}(c) shows the trajectory of UAVs, which reveals that UAVs can dynamically balance data collection in non-charging zones with strategic returns to LBDs for recharging, avoiding both energy depletion and excessive data staleness.

\section{Future Directions}
\label{Section: Future Directions}

{\color{color_blue}{
\subsection{Miniaturization of High-Efficiency Photovoltaic Receivers}
\label{Subsection: Miniaturization of High-Efficiency Photovoltaic Receivers}

\par The widespread adoption of WLPT in resource-constrained IoT endpoints demands the development of compact, lightweight and highly efficient photovoltaic receivers. Current receiver modules are often bulky and inefficient due to limitations in material design, optical coupling architecture, and thermal management.  Specifically, conventional semiconductor-based PV cells exhibit suboptimal responsivity at laser wavelengths and suffer from heat-induced performance degradation under high optical intensities.  To overcome these challenges, future efforts should focus on the co-design of photonic and electronic structures to achieve efficient photon-to-electricity conversion with minimal losses.  Promising directions include the use of meta-material absorbers to enhance light trapping and spectral selectivity, tandem multi-junction cells engineered for wavelength matching with specific laser sources, and integrated micro-optical concentrators that improve beam alignment tolerance.  Moreover, incorporating advanced thermal dissipation layers and transparent conductive coatings could further improve both the conversion efficiency and operational stability of receiver modules, enabling scalable deployment of WLPT-powered IoT ecosystems.

\subsection{Adaptive Beam Tracking with Environmental Resilience}
\label{Subsection: Adaptive Beam Tracking with Environmental Resilience}

\par Reliable operation of WLPT in real-world environments requires robust beam tracking and stabilization capabilities that can dynamically compensate for atmospheric disturbances and platform motion.  Current systems often rely heavily on ideal line-of-sight conditions and static alignment mechanisms, which are inadequate for mobile or low-altitude UAV operations subject to wind-induced vibrations, thermal blooming, beam wander, and particulate scattering.  These factors can cause significant power fluctuations and alignment loss, thereby reducing transfer efficiency and system reliability.  To address these challenges, future research must establish hybrid sensing and control architectures that integrate real-time environmental perception and adaptive correction. Promising strategies include fusing multi-modal data from infrared thermography, lidar-based atmospheric profiling, and optical scintillation sensors to estimate beam distortion characteristics.  Coupled with predictive control algorithms and fast beamsteering optics such as micro-electro-mechanical mirrors or deformable lenses, these approaches could enable self-correcting WLPT links resilient to dynamic outdoor conditions.

\subsection{Integration of WLPT and Next-Generation Communication Networks}
\label{Subsection: Integration of WLPT and Next-Generation Communication Networks}

\par  The convergence of WLPT with next-generation communication networks such as 5G-Advanced and 6G enables joint optimization of power and data resources. Both technologies rely on ultra-low latency control and precise localization, making real-time coordination feasible. Future systems can exploit some network information, such as the position, battery level, and mission profile of low-altitude UAVs, to dynamically adjust WLPT beam targeting and power allocation, thereby achieving predictive and energy-efficient operation. Conversely, WLPT availability can assist network planning and scheduling, enabling energy-aware routing and mission timing. This integration transforms WLPT from a standalone power link into a core component of the 6G energy-information ecosystem, requiring standardized interfaces and coordination protocols for seamless interaction. Ultimately, such synergy will support truly autonomous and self-sustaining intelligent networks that unify communication and energy delivery.

\subsection{Safety Regulation Frameworks for WLPT}
\label{Subsection: Safety Regulation Frameworks for WLPT}

\par WLPT systems involve high-intensity optical beams, which pose potential hazards to human eyes, skin, and surrounding equipment. Consequently, robust safety regulation frameworks are essential to ensure operational compliance and risk mitigation. Existing guidelines provide baseline exposure limits and classification schemes for laser systems. However, these standards were primarily developed for stationary or industrial laser applications and may not fully capture the unique challenges of mobile or UAV-based WLPT deployments. Future regulatory efforts must account for dynamic operational scenarios, including moving platforms, varying atmospheric conditions, and multi-beam interference. This requires the development of adaptive safety protocols that integrate real-time beam power monitoring, automatic shutdown mechanisms, and predictive risk assessment algorithms. Additionally, certification processes should consider not only nominal beam intensity but also potential failure modes, alignment errors, and cumulative exposure effects over time. Establishing such comprehensive frameworks will be crucial for the large-scale adoption of WLPT in civilian and commercial IoT ecosystems while ensuring user safety and regulatory compliance.
}}
\section{Conclusion}
\label{Section: Conclusion}

\par This paper has investigated WLPT as an efficient solution for energy-constrained low-altitude UAV-assisted IoT networks. We have systematically analyzed principles and advantages of WLPT over conventional wireless charging methods. Three operational paradigms have been proposed for system integration. Moreover, we have identified some key challenges in WLPT for low-altitude UAV-assisted and discussed some potential solutions. Our case study has illustrated that the proposed MAPPO-TM framework significantly enhances energy sustainability and data freshness in WLPT for UAV-assisted data collection scenarios. Furthermore, we have outlined some important future directions.

\ifCLASSOPTIONcaptionsoff
\newpage
\fi

\bibliographystyle{IEEEtran}
\bibliography{mybib}

@Article{Wang2025,
  author  = {Wang, Yixian and Sun, Geng and Sun, Zemin and Wang, Jiacheng and Li, Jiahui and Zhao, Changyuan and Wu, Jing and Liang, Shuang and Yin, Minghao and Wang, Pengfei and Niyato, Dusit and Sun, Sumei and Kim, Dong In},
  journal = {{IEEE} Trans. Cognit. Commun. Networking},
  title   = {Toward Realization of Low-Altitude Economy Networks: Core Architecture, Integrated Technologies, and Future Directions},
  year    = {Early Access, 2025},
  note    = {doi: {10.1109/TCCN.2025.3601015}},
}

@Article{He2024,
  author    = {Bin He and Xiangxin Ji and Gang Li and Bin Cheng},
  journal   = {{IEEE} Trans. Cogn. Commun. Netw.},
  title     = {Key Technologies and Applications of {UAVs} in Underground Space: {A} Review},
  year      = {2024},
  month     = {Mar.},
  number    = {3},
  pages     = {1026--1049},
  volume    = {10},
}

@Article{Yuan2025,
  author        = {Weijie Yuan and Yuanhao Cui and Jiacheng Wang and Fan Liu and Geng Sun and Tao Xiang and Jie Xu and Shi Jin and Dusit Niyato and Sinem Coleri and Sumei Sun and Shiwen Mao and Abbas Jamalipour and Dong In Kim and Mohamed-Slim Alouini and Xuemin Shen},
  title         = {From Ground to Sky: Architectures, Applications, and Challenges Shaping Low-Altitude Wireless Networks},
  year          = {2025},
  month         = {Jun.},
  journal = {arXiv preprint arXiv:2506.12308v2},
  note = {doi: {10.48550/arXiv.2506.12308}},
}

@Article{Liu2025,
  author  = {Lingling Liu and Aimin Wang and Geng Sun and Jiahui Li and Hongyang Pan and Tony Q. S. Quek},
  journal = {{IEEE} Trans. Veh. Technol.},
  title   = {Multi-Objective Optimization for Data Collection in {UAV}-Assisted Agricultural {IoT}},
  year    = {2025},
  month   = {Apr.},
  number  = {4},
  pages   = {6488--6503},
  volume  = {74},
}

@Article{Gupta2025,
  author    = {Ajay Kumar Gupta and Manav R. Bhatnagar},
  journal   = {{IEEE} Trans. Green Commun. Netw.},
  title     = {A Non-Cooperative Pricing Strategy for {UAV}-Enabled Charging of Wireless Sensor Network},
  year      = {2025},
  month     = {Feb},
  number    = {2},
  pages     = {459--470},
  volume    = {9},
}

@Article{Yan2020,
  author    = {Hua Yan and Yunfei Chen and Shuang{-}Hua Yang},
  journal   = {{IEEE} Trans. Veh. Technol.},
  title     = {{UAV}-Enabled Wireless Power Transfer With Base Station Charging and {UAV} Power Consumption},
  year      = {2020},
  month     = {Nov.},
  number    = {11},
  pages     = {12883--12896},
  volume    = {69},
}

@Article{Hassija2020,
  author    = {Vikas Hassija and Vinay Chamola and Dara Nanda Gopala Krishna and Mohsen Guizani},
  journal   = {{IEEE} Trans. Veh. Technol.},
  title     = {A Distributed Framework for Energy Trading Between UAVs and Charging Stations for Critical Applications},
  year      = {2020},
  month     = {May},
  number    = {5},
  pages     = {5391--5402},
  volume    = {69},
}

@Article{Ren2025,
  author    = {Meixuan Ren and Haipeng Dai and Tang Liu and Xianjun Deng and Wanchun Dou and Yuanyuan Yang and Guihai Chen},
  journal   = {{IEEE} Commun. Surv. Tutorials},
  title     = {Understanding Wireless Charger Networks: Concepts, Current Research, and Future Directions},
  year      = {2025},
  month     = {Aug.},
  number    = {4},
  pages     = {2247--2282},
  volume    = {27},
}

@Article{Ahmadi2025,
  author    = {Kimia Ahmadi and Wouter A. Serdijn},
  journal   = {{IEEE} Internet Things J.},
  title     = {Advancements in Laser and {LED}-Based Optical Wireless Power Transfer for {IoT} Applications: {A} Comprehensive Review},
  year      = {2025},
  month     = {Dec.},
  number    = {12},
  pages     = {18887--18907},
  volume    = {12},
}

@Article{Wang2023,
  author={Wang, Yao and Zhang, Hua and Cao, Yue and Lu, Fei},
  journal={{IEEE} Trans. Transp. Electrif.}, 
  title={Remaining Opportunities in Capacitive Power Transfer Based on Duality With Inductive Power Transfer}, 
  year={2023},
  month={Jun.},
  volume={9},
  number={2},
  pages={2902-2915},
}

@Article{Xie2025,
  author  = {Wenwen Xie and Geng Sun and Jiahui Li and Jiacheng Wang and Hongyang Du and Dusit Niyato and Octavia A. Dobre},
  journal = {{IEEE} Internet Things Mag.},
  title   = {Generative {AI} for Energy Harvesting {Internet} of Things Network: Fundamental, Applications, and Opportunities},
  year    = {2025},
  month   = {May},
  number  = {3},
  pages   = {72--80},
  volume  = {8},
}

@ARTICLE{Jin2019,
  author={Jin, Ke and Zhou, Weiyang},
  journal={{IEEE} Trans. Power Electron.}, 
  title={Wireless Laser Power Transmission: A Review of Recent Progress}, 
  year={2019},
  month={Apr.},
  volume={34},
  number={4},
  pages={3842-3859},
}

@InProceedings{Xue2025,
  author    = {Chaocan Xue and Bineng Zhong and Qihua Liang and Yaozong Zheng and Ning Li and Yuanliang Xue and Shuxiang Song},
  booktitle = {Proc. {IEEE} Conf. Comput. Vis. Pattern Recognit. (CVPR)},
  title     = {Similarity-Guided Layer-Adaptive Vision Transformer for {UAV} Tracking},
  year      = {2025},
  month     = {Jun. 11-15,},
  address   = {Nashville, TN, USA},
  pages     = {6730--6740},
}

@Article{Dou2025,
  author  = {Jizhe Dou and Haotian Zhang and Yang Luo and Guodong Sun},
  journal = {{IEEE} Trans. Mob. Comput.},
  title   = {Scheduling Drone and Mobile Charger via Hybrid-Action Deep Reinforcement Learning},
  year    = {2025},
  month   = {Aug.},
  number  = {8},
  pages   = {7788--7804},
  volume  = {24},
}

@Article{Zhang2025,
  author        = {Wen Zhang and Aimin Wang and Jiahui Li and Geng Sun and Jiacheng Wang and Weijie Yuan and Dusit Niyato},
  title         = {Energy Transfer and Data Collection from Batteryless Sensors in Low-altitude Wireless Networks},
  year          = {2025},
  month         = {Jul.},
  journal = {arXiv preprint arXiv:2507.07481},
  note = {doi: {10.48550/ArXiv.2507.07481}},
}
\vfill

\end{document}